\documentclass[a4paper,twocolumn,showpacs,superscriptaddress,floatfix,nofootinbib]{revtex4-1}
\usepackage{graphicx}
\usepackage{epsf}
\usepackage{epsfig}
\usepackage{amssymb,amsmath}
\usepackage[usenames]{color}
\usepackage{amssymb}
\usepackage{times}
\usepackage{hyperref}
\hypersetup{
  colorlinks=true,        % false: boxed links; true: colored links
  linkcolor=blue,         % color of internal links
  citecolor=cyan,         % color of links to bibliography
}

\newcommand{\cf}{cf.~}
\newcommand{\ie}{i.e.,~}
\newcommand{\eg}{e.g.,~}
\renewcommand{\BibitemShut}[1]{}
\begin{document}

\title{Constraining the Equation of State of Neutron Stars from Binary
  Mergers}
%%%
\author{Kentaro~Takami}
\affiliation{Institut f{\"u}r Theoretische Physik, Max-von-Laue-Stra{\ss}e 1, 60438
Frankfurt, Germany}
%%%
\author{Luciano~Rezzolla}
\affiliation{Institut f{\"u}r Theoretische Physik, Max-von-Laue-Stra{\ss}e  1, 60438
Frankfurt, Germany}
\affiliation{Max-Planck-Institut f{\"u}r Gravitationsphysik, Albert Einstein
Institut, Am M{\"u}hlenberg 1, 14476 Potsdam, Germany}
%%%
\author{Luca~Baiotti}
\affiliation{Graduate School of Science and Institute of Laser
Engineering, Osaka University, Toyonaka 5600043, Japan}
%%%

\begin{abstract}
Determining the equation of state of matter at nuclear density and hence
the structure of neutron stars has been a riddle for decades. We show how
the imminent detection of gravitational waves from merging neutron star
binaries can be used to solve this riddle. Using a large number of
accurate numerical-relativity simulations of binaries with nuclear
equations of state, we find that the postmerger emission is
characterized by two distinct and robust spectral features. While the
high-frequency peak has already been associated with the oscillations of
the hypermassive neutron star produced by the merger and depends on the
equation of state, a new correlation emerges between the low-frequency
peak, related to the merger process, and the total compactness of the
stars in the binary. More importantly, such a correlation is essentially
universal, thus providing a powerful tool to set tight constraints on the
equation of state. If the mass of the binary is known from the inspiral
signal, the combined use of the two frequency peaks sets four
simultaneous constraints to be satisfied. Ideally, even a single
detection would be sufficient to select one equation of state over the
others. We test our approach with simulated data and verify it
works well for all the equations of state considered.
\end{abstract}

\pacs{
04.25.Dm, % numerical relativity
04.25.dk,  %Numerical studies of other relativistic binaries
% 04.25.Nx,  % Post-Newtonian approximation; perturbation theory; related approximations
04.30.Db, % gravitational wave generation and sources
04.40.Dg, % Relativistic stars: structure, stability, and oscillations
% 04.70.Bw, % classical black holes
95.30.Lz, % Hydrodynamics
95.30.Sf, % relativity and gravitation
97.60.Jd % Neutron stars
% 97.60.Lf  % black holes (astrophysics)
}

\maketitle

%%%%%%%%%%%%%%%%%%%%%%%%%%%%%%%%%%%%%%%%%%%%%%%%%%%%%%%%%%%%%%%%%
%%%   MAIN TEXT
%%%%%%%%%%%%%%%%%%%%%%%%%%%%%%%%%%%%%%%%%%%%%%%%%%%%%%%%%%%%%%%%%

\emph{Introduction.~~} This decade is likely to witness the first direct
detection of gravitational waves (GWs) as a series of advanced detectors
such as LIGO~\citep{Harry2010}, Virgo~\citep{Accadia2011_etal}, and
KAGRA~\citep{Aso:2013} become operational in the next five years.  Among
the sources of GWs expected to be detected are the inspiral and
postmerger of neutron-star binaries or neutron-star--black-hole binaries,
and binary black holes. Population-synthesis models suggest that binary
neutron star mergers (BNSs) may be the most common source, with an
expected detection rate of ${\sim}40\,\mathrm{yr}^{-1}$~\cite{Abadie:2010_etal}.

Any GW signal from a binary including a neutron star will contain
important signatures of the equation of state (EOS) of matter at nuclear
densities. A first signature is represented by the tidal corrections to
the orbital phase; these are reasonably well understood
analytically~\cite{Flanagan08, Baiotti:2010, Bernuzzi2012} and can be
tracked accurately with advanced high-order numerical
codes~\cite{Radice2013b,Radice2013c}. A second signature is instead
related to the postmerger phase, where the object formed by the merger
 [most likely a hypermassive neutron star (HMNS)] can emit GWs in a
narrow frequency range before collapsing to a black
hole~\cite{Baiotti08}.

The first evidence that the information contained in the postmerger
signal could be extracted from the corresponding spectrum was provided by
Bauswein and Janka~\cite{Bauswein2011} (see
also Refs.~\cite{Oechslin07b,Bauswein2012}), who performed a large number of
simulations using a smoothed particle hydrodynamics code solving the
Einstein field equations assuming conformal flatness, and employing a GW
backreaction scheme within a post-Newtonian approximation (see
also Ref.~\cite{Hotokezaka2013c} for a subsequent general-relativistic
study). Reference~\cite{Bauswein2011}, in particular, pointed out the presence
of a peak at high-frequency in the spectrum (dubbed $f_{\rm peak}$) and
showed it correlated with the properties of the EOS, \eg with the radius
of the maximum-mass nonrotating configuration. It was then recognized
that $f_{\rm peak}$ corresponds to a fundamental fluid mode with $m=2$ 
of the HMNS~\cite{Stergioulas2011b}.

By performing a large number of accurate simulations in full general
relativity of equal-mass and unequal-mass BNSs with a number of different
nuclear EOSs, we have revisited the spectral properties of the postmerger
GW signal. In this Letter we report our analysis of the spectral
features with special attention to the low-frequency peak, which tracks
the strong emission produced at the merger when the two dense stellar
cores collide. We show that this peak has an essentially universal
relation with the total compactness of the stars in the binary so that,
combining the information from the two peaks, we can derive a simple and
robust method to constrain the EOS.

%%%%%%%%%%%%%%%%%%%%%%%%%%%%%%

\emph{Numerical Setup.~~} Our results have been obtained in full general
relativity solving the Einstein equations with the \texttt{McLachlan}
code \cite{Brown:2008sb,Loffler:2011ay}. The solution of the relativistic
hydrodynamics equations is instead obtained using the \texttt{Whisky}
code~\citep{Baiotti04,Baiotti08}. The stars are modeled as obeying a
nuclear EOS and we have considered five different models: \ie
APR4~\cite{Akmal1998a}, ALF2~\cite{Alford2005}, SLy~\cite{Douchin01},
H4~\cite{GlendenningMoszkowski91}, GNH3~\cite{Glendenning1985}. Rather than
using tables, it is more convenient to use $n$ piecewise polytropic
approximations to these EOSs~\cite{Read:2009a}, expressing the ``cold''
contribution to pressure and specific internal energy as $p_\mathrm{c} =
K_i \rho^{\Gamma_i}$, $\epsilon_\mathrm{c} = \epsilon_i + K_i\rho
^{\Gamma_i-1}/(\Gamma_i -1)$, where $\rho$ and $K$ are the rest-mass
density and the polytropic constant, respectively
(see Ref.~\cite{Rezzolla_book:2013} for details); $n=4$ is sufficient to
obtain a rather accurate representation of the different EOSs.

In addition, to model the thermal effects arising from the merger, the
cold pressure is augmented through an ideal-fluid EOS, so that the total
pressure and specific internal energy are $p = p_\mathrm{c} +
p_\mathrm{th}$, $\epsilon = \epsilon_\mathrm{c} + \epsilon_\mathrm{th}$,
with $p_\mathrm{th} = \rho \epsilon_\mathrm{th} \left(\Gamma_\mathrm{th}
-1 \right)$~\cite{Janka93}.  Following Ref.~\cite{Bauswein:2010dn}, we use
$\Gamma_\mathrm{th}=2$, but we have verified that our results are not
sensitive to this choice, with spectral differences that are a few
percent at most when, for instance, $\Gamma_\mathrm{th}=1.8$ (a full
analysis will be reported in a longer paper~\cite{Takami:2014}). Finally,
to span a larger range in stellar compactness and go beyond the one
covered by the nuclear EOSs above, we have considered a sixth EOS given
by a pure polytrope with $\Gamma=2$ and $K=123.6$ in units where $c=G=M_\odot=1$.

For each EOS we have considered five equal-mass binaries with average
(gravitational) mass at infinite separation in the range $\bar{M} \equiv
(M_1+M_2)/2 = (1.275-1.375)M_{\odot}$ for the APR4 EOS,
$(1.225-1.325)M_{\odot}$ for the ALF2 EOS, $(1.250-1.350)M_{\odot}$ for
the GNH3, H4, and SLy EOSs, and $(1.350-1.450)M_{\odot}$ for the
$\Gamma=2$ polytrope (higher masses would lead to short-lived HMNSs;
additional properties of the binaries will be presented in the longer
paper). We have also considered two unequal-mass binaries for the
GNH3 and SLy EOSs having $\bar{M}=1.300\,M_{\odot}$ and mass ratio ${\simeq}
0.92$. The binaries are modeled as irrotational in quasicircular orbits
and computed with the \texttt{LORENE}
code~\citep{Gourgoulhon-etal-2000:2ns-initial-data}, assuming a
conformally flat metric. To increase resolution we have employed a
reflection symmetry across the $z=0$ plane, a $\pi$-symmetry condition
across the $x=0$ plane (only for equal-mass binaries), and a moving-mesh
refinement via the \texttt{Carpet} driver~\citep{Schnetter-etal-03b}. We
have used six refinement levels, the finest having a resolution of
$0.15\,M_{\odot} \simeq 0.221\,\mathrm{km}$, and extracted the GWs near
the outer boundary at a distance $R_{0}=500\,M_{\odot} \simeq
738\,\mathrm{km}$.

%%%%%%%%%%%%%%%%%%%%%%%%%%%%%%

\begin{figure}
\centering
\includegraphics[width=0.95\columnwidth]{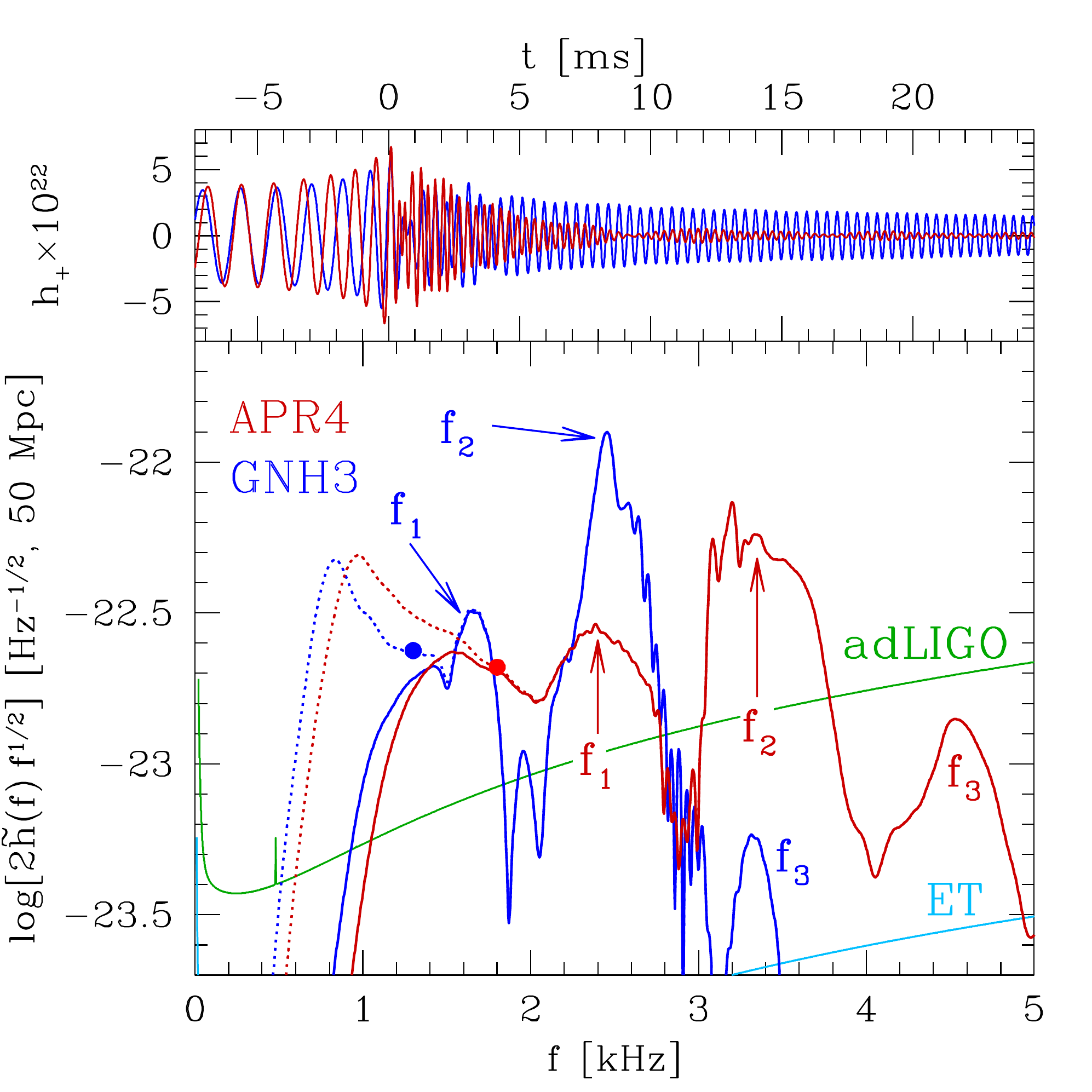}
\caption{\label{fig:1} Top panel: evolution of $h_+$ for
  representative binaries with the APR4 and GNH3 EOSs (dark-red and blue
  lines, respectively) for sources at a polar distance of 50
  Mpc. Bottom panel: spectral density $2 \tilde{h}(f) f^{1/2}$
  windowed after the merger for the two EOSs and sensitivity curves of
  Advanced LIGO (green line) and ET (light-blue line); the dotted lines
  show the power in the inspiral, while the circles mark the contact
  frequency.}
\end{figure}

\emph{Results.~~} As discussed by several authors~\cite{Oechslin07b,
  Baiotti08, Bauswein2011, Bauswein2012, Hotokezaka2013c, Messenger2013},
the power spectral density (PSD) of the postmerger GW signal exhibits a
number of clear peaks. Two examples are presented in Fig.~\ref{fig:1},
which refers to two binaries with (gravitational) masses
$\bar{M}/M_{\odot}=1.325$, and APR4 and GNH3 EOSs. 
Since $h_+ \sim h_{+}^{22}$, the top panel shows the evolution of the
$\ell=m=2$ plus polarization of
the strain aligned at the merger~\cite{Read2013} for sources at a
polar distance of 50 Mpc (dark-red and blue lines for the APR4 and GNH3
EOSs, respectively). The bottom panel, on the other hand, shows the
spectral densities $2 \tilde{h}(f) f^{1/2}$ windowed after the merger for
the two EOSs, comparing them with the sensitivity curves of Advanced
LIGO~\cite{url:adLIGO_Sh_curve} (green line) and of the Einstein
Telescope~\cite{Punturo2010b,Sathyaprakash:2009xs} (ET; light-blue
line). The dotted lines refer to the whole time series and hence, where
visible, indicate the power during the inspiral, while the circles mark
the ``contact frequency'' $f_{\rm cont}=\mathcal{C}^{3/2}/(2\pi
\bar{M})$~\cite{Damour:2012}, where $\mathcal{C} \equiv \bar{M}/\bar{R}$
is the average compactness, $\bar{R} \equiv (R_1+R_2)/2$, and $R_{1,2}$
are the radii of the nonrotating stars associated with each binary.

Note the clear appearance of two main peaks, indicated as $f_1$ and
$f_2$, with the first one being smaller in amplitude, but also at
frequencies where the detectors are more sensitive. The $f_2$ peak was
named $f_{\rm peak}$ in Refs.~\cite{Bauswein2011,Bauswein2012} and $f_2$
in Ref.~\cite{Stergioulas2011b}. When comparing our values of $f_2$ with the
corresponding ones from Ref.~\cite{Bauswein2011} for the same binaries, we
have found differences of the order of a few percent at most, thus
confirming that the conformally flat approximation provides a rather
accurate description of the dynamics of the HMNS. The amplitude of the GW
emission, on the other hand, is ${\sim}6$--$9$ times larger than
in Ref.~\cite{Bauswein2011}.  We should also note that
Ref.~\cite{Stergioulas2011b} reported the presence of two additional
frequencies, dubbed $f_{-} < f_2$ and $f_{+} > f_2$, where $f_{-}$ was
tentatively attributed to a nonlinear interaction between the quadrupole
and quasiradial modes. We share this interpretation and, as already done
in Ref.~\cite{Baiotti08}, recognize $f_1$ (and thus $f_{-}$
of Ref.~\cite{Stergioulas2011b}) as produced by the nonlinear oscillations of
the two stellar cores that collide and bounce repeatedly right after the
merger. A larger uncertainty is associated with the physical
interpretation of the third and very high-frequency peak, which is
indicated as $f_3$ in Fig.~\ref{fig:1}.  Additional work is needed to
explain this mode, which could be an overtone or the result of the
nonlinear interaction of the $f_2$ mode with other nonquasiradial modes.

Although clearly recognizable, we have preferred to use an automatic
evaluation of the peak frequencies using a prescription similar to the
one discussed in Ref.~\cite{Messenger2013}, where a fit of the PSDs is performed 
using two different Gaussian profiles for the two peaks. Details on
the fitting procedure of the PSD and the associated errors will be
presented in the longer paper~\cite{Takami:2014}.

\begin{figure}
\centering
\includegraphics[width=\columnwidth]{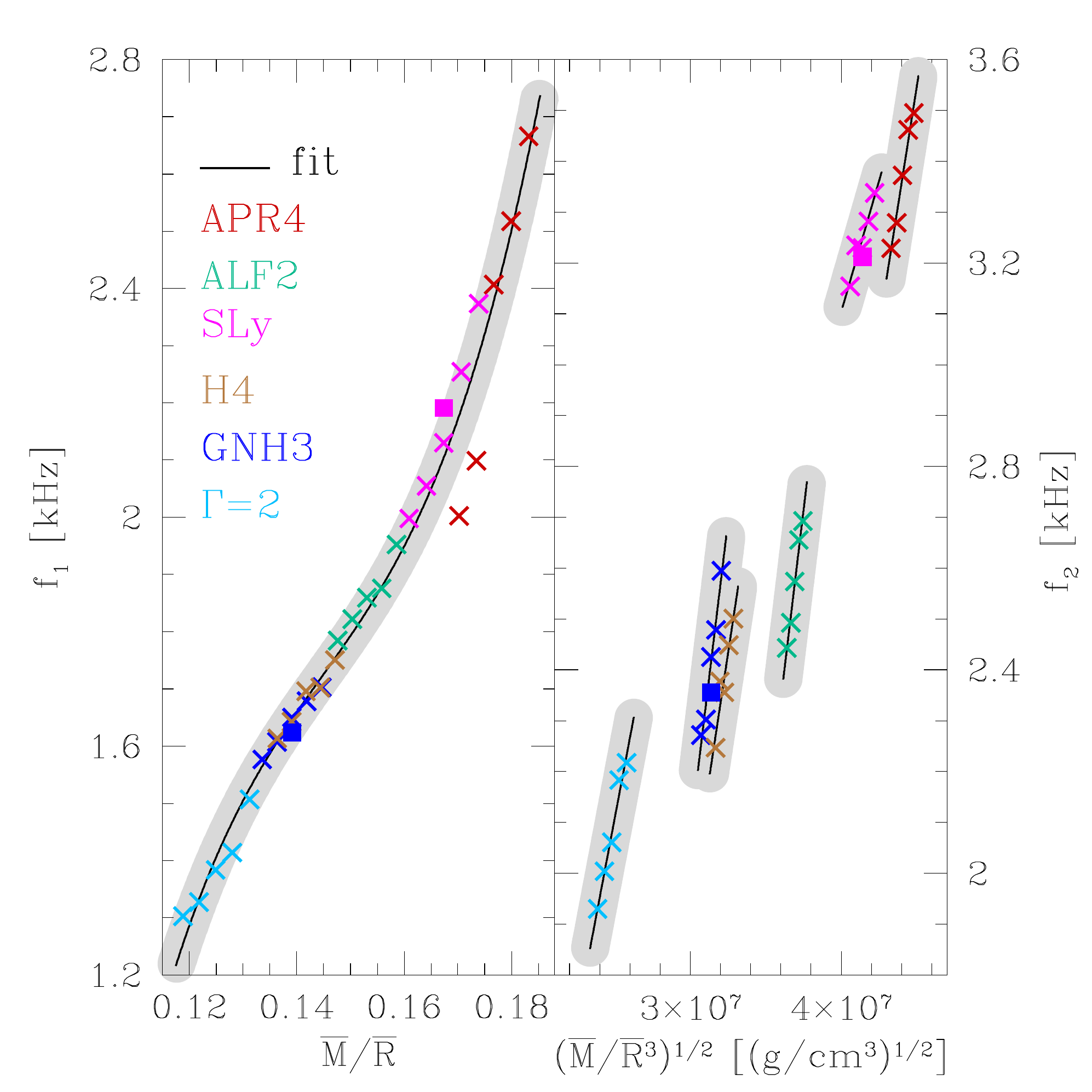}
\caption{\label{fig:2} Left panel: fitted values of the
  low-frequency peaks as a function of the stellar compactness for the
  six different EOSs considered; note the universal behavior exhibited
  also by unequal-mass binaries (filled squares). Shown as a solid black
  line is the cubic fit, while the gray band is the estimate of the total
  errors. Right panel: fitted values of the high-frequency peaks
  as a function of the average rest-mass density; no universal behavior
  appears.}
\end{figure}

The most interesting and important result of our spectral analysis is
that there is a very clear correlation between the low-frequency peak
$f_1$ and the compactness $\mathcal{C}$. The results of the fitting
procedure for the low-frequency peak are collected in the left panel of
Fig.~\ref{fig:2}, where the values of the fitted $f_1$ frequencies are
plotted as a function of $\mathcal{C}$ for the various EOSs and are
indicated with different colors. Note that the plot displays
high-accuracy data for 32 BNSs (the unequal-mass binaries appear as
filled squares) and thus collects the results of one year of computing
time. Also shown as a shaded gray band is the estimate of the total
error, which is effectively dominated by the fitting procedure of the
PSD. Postponing to the longer paper the details of the error budget, we
anticipate here that the average numerical error from the simulations is
$0.06$ kHz, while the average uncertainty in the fitting procedure of the
PSD is of $0.2$ kHz (see also Ref.~\cite{Messenger2013}).

\begin{figure*}
\centering
%\includegraphics[width=0.49\columnwidth]{./Constr_ALF2_M13250.pdf}
%\hskip 0.0cm
%\includegraphics[width=0.49\columnwidth]{./Constr_APR4_M13250.pdf}
\includegraphics[width=0.95\columnwidth]{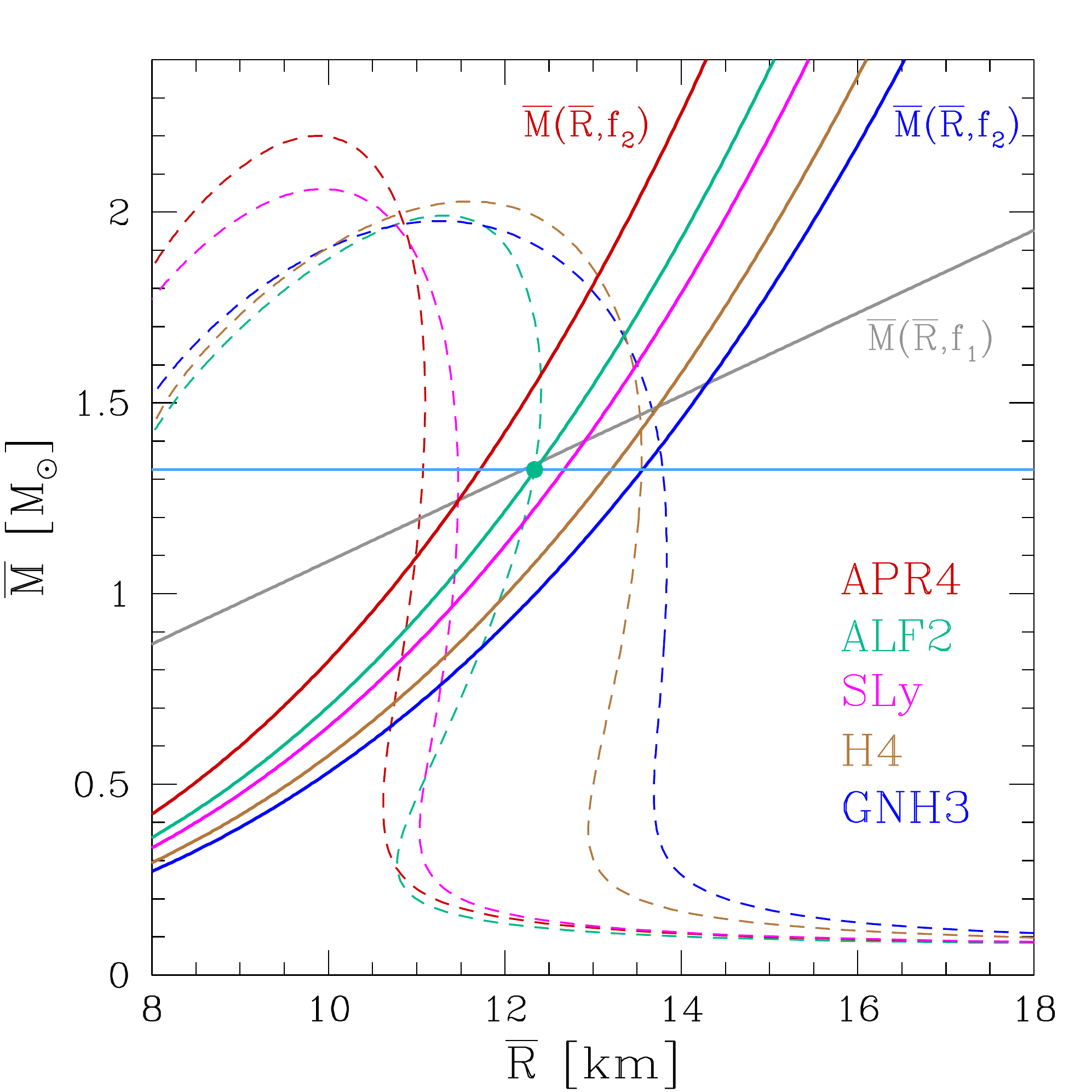}
\hskip 0.5cm
\includegraphics[width=0.95\columnwidth]{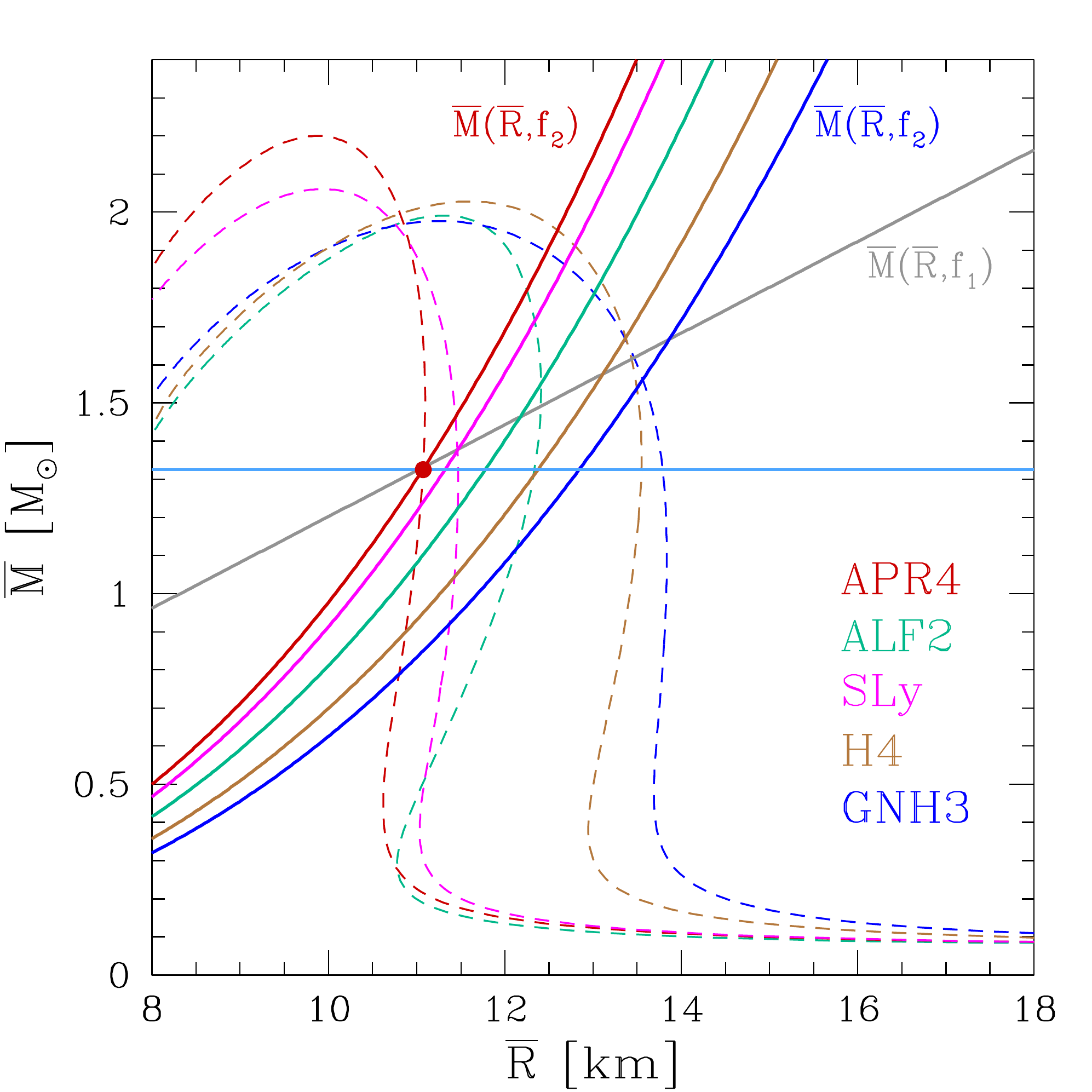}
\caption{\label{fig:3} Examples of use of the spectral features to
  constrain the EOS. Once a detection is made, the relations
  $\bar{M}=\bar{M}(\bar{R},f_1)$ and
  $\bar{M}=\bar{M}(\bar{R},f_2;\mathrm{EOS})$ (colored solid lines) will
  cross at one point the curves of equilibrium configurations (colored
  dashed lines). Knowledge of the mass of the system (horizontal line)
  will provide a fourth constraint, removing possible degeneracies. The
  left and right panels refer to the ALF2 and APR4 EOSs, but all other
  EOSs behave in the same way.}
\end{figure*}

The behavior of the low-frequency peak is remarkably consistent with a
simple polynomial function and we have found that a cubic polynomial
provides the best fit (solid black line). In this case, the chi-squared
value measured is $0.09$, with a fitting uncertainty ${\lesssim}0.06$~kHz; even
if the data relative to the APR4 EOS have the largest scattering, all
simulations are very well reproduced by the fit within the error
bars. The essentially universal behavior of the $f_1$ frequency with
compactness is reminiscent of another universal behavior shown by the
orbital frequency at the peak of the GW
amplitude~\cite{Read2013,Bernuzzi2014} and provides a powerful tool to
constrain the EOS. This is because once a measurement of $f_1$ is made,
the fitting provides a relation of the type
$\bar{M}=\bar{M}(\bar{R},f_1)$, which intersects in just one point in the
$(\bar{M},\bar{R})$ plane the relation $\bar{M}=\bar{M}(\bar{R})$ built
for each EOS through equilibrium nonrotating models (\cf Fig.~\ref{fig:3}
and see discussion below).

Shown instead in the right panel of Fig.~\ref{fig:2} is the behavior of
the high-frequency peak $f_2$ as a function of the average rest-mass
density $\left(\bar{M}/\bar{R}^3\right)^{1/2}$ for the different
EOSs. 
(A correlation with $\mathcal{C}$ is present also for the
$f_2$ frequency, but with a much larger scatter, making it difficult to
use to obtain robust and independent information. Correlations are
also possible between $f_2$ and other quantities such as the stellar
radius at fixed mass~\cite{Bauswein2011,Bauswein2012,Hotokezaka2013c},
but again with a large scatter and strong dependence on the EOS.)
A similar plot of $f_2$ as a function of
$(\bar{M}/R^3_{\mathrm{max}})^{1/2}$, where $R_{\mathrm{max}}$ is the
radius of the maximum-mass nonrotating configuration, was presented in
Ref.~\cite{Bauswein2011}, where only one mass was considered for the
different binaries, but a larger set of EOSs was used. Overall, the mass
dependence in $f_2$ (\ie what distinguishes different points of the same
color) does not suggest a tight universal correlation in our data. Hence,
we perform a linear fit for each EOS, reproducing the data rather well
(the chi-squared value is ${\lesssim}0.004$). Although EOS dependent, these fits
still provide a set of relations
$\bar{M}=\bar{M}(\bar{R},f_2;\mathrm{EOS})$, which will again intersect
at just one point the sequences of equilibrium nonrotating models for
each EOS (\cf Fig.~\ref{fig:3}).

Armed with the correlations described above, we now discuss how to use
them to constrain the EOS. Let us assume that the GW signal from a BNS
has been detected and that the source is sufficiently close that all of
the spectral features are clearly identifiable.
[A ``clear identification'' will need a high signal-to-noise (SNR) ratio and will
depend on the EOS; for those in Fig.~\ref{fig:1}, a realistic
$\mathrm{SNR}=5$ of Advanced LIGO for $f_1$ implies sources at
distances of ${\sim}25({\sim}40)$ Mpc for the APR4(GNH3) EOSs; large distances
  of ${\sim}50({\sim}115)$ Mpc are possible for $f_2$.]
Using the measured
values of $f_1$ and $f_2$ we can draw on the $(\bar{M},\bar{R})$ plane a
series of curves given by the relations $\bar{M}=\bar{M}(\bar{R},f_1)$
(solid gray line) and $\bar{M}=\bar{M}(\bar{R},f_2;\mathrm{EOS})$ (solid
colored lines). This is shown in Fig.~\ref{fig:3}, where the left panel
refers to the ALF2 EOS, while the right one refers to the APR4
EOS. Concentrating on the left panel, we can see that the
$\bar{M}=\bar{M}(\bar{R},f_1)$ relation intersects each of the various
equilibrium curves (colored dashed lines) at one point (\eg at $\bar{M}
\simeq 1.325\,M_{\odot},\,\bar{R} \simeq 12.3$ km for the ALF2 EOS), but
also other crossings take place for the other EOSs. However, when using
also the relations $\bar{M}=\bar{M}(\bar{R},f_2;\mathrm{EOS})$, some
EOSs can be readily excluded (\eg APR4, SLy, and GNH3, in our example) and
only the ALF2 and H4 EOSs have crossings (or ``near crossings'') between
the equilibrium-models curves and the frequency-correlations
curves. Realistically, the uncertainties in the measurement of $f_{1,2}$
(including the experimental ones) will make the correlation curves
$\bar{M}=\bar{M}(\bar{R},f_1)$ and
$\bar{M}=\bar{M}(\bar{R},f_2;\mathrm{EOS})$ appear as ``bands'' with
probability distributions rather than thin lines; the crossing will be
harder to judge and will require a complete Bayesian probability analysis
(see, \eg Ref.~\cite{Messenger2013}), which is beyond the scope of this Letter;
hence, by near crossings we here mean the overlap of different curves
in a small region of the $(\bar{M},\bar{R})$ plane.

Fortunately, the uncertainty can be removed if the mass of the binary is
known from the inspiral signal. In this case, in fact, there will be a
horizontal line in the $(\bar{M},\bar{R})$ plane that will break the
degeneracy imposing four simultaneous constraints. This is shown with the
horizontal light-blue line, which clearly intersects the three curves
relative to the ALF2 EOS at one point only (green solid circle). Also in
this case, the horizontal line should in reality be replaced by a band
with a probability distribution, but from Fig.~\ref{fig:2} it is already
possible to conclude that the mass needs to be determined with a relative
precision that is ${\lesssim}10\%$.

Despite the simplifying assumptions, this method shows that even a single
detection of a GW signal with high SNR and from which the mass of the
binary can be calculated, would be sufficient to set tight constraints on
the EOS. This approach works well for all of the binaries considered and
an additional example is offered by the right panel of Fig.~\ref{fig:3},
which reproduces a similar construction for the APR4 EOS. Clearly, also
in this case four different curves cross essentially at one point.

Of course this method can work as long as there is a sufficient number of
detections and the uncertainties in the measure of the frequencies are
small. Using the postmerger signal and fixing a realistic
$\mathrm{SNR}=5$, different EOSs and optimally oriented binaries yield a
detection horizon of ${\sim}20$--$40$ Mpc, which reduces to ${\sim}14$--$28$ Mpc 
for randomly oriented sources. In turn, the latter yields
an event rate of ${\sim}0.01$--$0.1\,\mathrm{yr}^{-1}$, which could increase to
${\sim}0.1$--$1\,\mathrm{yr}^{-1}$ for the optimistic estimate
of Ref.~\cite{Abadie:2010_etal}. We note that if we assume $\mathrm{SNR}=2$ as
in~ Ref.~\cite{Bauswein2011}, then our expected event rate is larger by a
factor of $(5/2)^3 \simeq 16$. Following Refs.~\cite{Read:2009b}
and~\cite{Bauswein2012}, we have used the Fisher information matrix to
estimate the uncertainties in the determination of the peak frequencies
when a GW detection is made. In particular, for sources with optimal
orientation at 50 Mpc, the uncertainties for adjacent models are in the
range ${\sim}1$--$100$ Hz, with the upper value being smaller than the one
reported in Ref.~\cite{Bauswein2011}, where distances of 20 Mpc were
considered.

A few additional remarks will be given before concluding. First, even if the
measurement of the mass is not available from the inspiral, the possible
degeneracies mentioned above could be removed with a few positive
detections, which would tend to favor one EOS over the others. Second,
if only the $f_2$ frequency is measurable, the approach discussed above
can still be used as long as the mass is known; in this case three and
not four curves will have to cross at one point. Third, most of our
simulations refer to equal-mass binaries, but we expect that
$f_{1,2}$ will not be very sensitive to the initial mass ratio; this was
already shown by Refs.~\cite{Bauswein2011,Hotokezaka2013c} and is confirmed by
the two unequal-mass binaries simulated. Fourth, realistic values of the
spins should not influence the frequencies significantly given that the
largest contribution to the angular momentum of the HMNS comes from the
orbital angular momentum and not from the initial spins of the
stars~\cite{Kastaun2013}. Finally, because the $f_1$ peak is produced
soon after the merger, it should not be affected significantly by
magnetic fields and radiative effects, whose modifications emerge on much
larger time scales~\cite{Giacomazzo:2010}.

%%%%%%%%%%%%%%%%%%%%%%%%%%%%%%
\emph{Conclusions.~~} We have carried out a large number of accurate and
fully general-relativistic simulations of the inspiral and postmerger of
BNSs with nuclear EOSs. This has allowed us to have a comprehensive view
of the spectral properties of the complex postmerger GW signal and to
highlight the presence of two robust frequency peaks. We have shown for
the first time that the low-frequency peaks exhibit a correlation with
the stellar compactness that is essentially EOS independent and that can
be used to constrain the EOS once the peak is measured. In addition, the
combined use of other EOS-dependent correlations from the high-frequency
peaks can further constrain the EOS. In principle, if the mass is known
from the inspiral and the peaks are clearly measurable, a single
detection would be sufficient to set constraints on the EOS. In practice,
a few detections will favor statistically one EOS over the others, but a
Bayesian analysis is necessary to quantify these probabilities; we leave
this to future work.

%%%%%%%%%%%%%%%%%%%%%%%%%%%%%%%%%%%%%%%%%%%%%%%%%%%%%%%%%%%%%%%%%
%%%   ACKNOWLEDGMENTS
%%%%%%%%%%%%%%%%%%%%%%%%%%%%%%%%%%%%%%%%%%%%%%%%%%%%%%%%%%%%%%%%%

\smallskip\noindent\emph{Acknowledgements.~~} We thank A. Bauswein,
H.-T. Janka, B. Sathyaprash, and N. Stergioulas for useful
comments. Partial support comes from the DFG grant SFB/Transregio~7 and
by ``NewCompStar'', COST Action MP1304. K.T. is supported by the
LOEWE-Program in HIC for FAIR. The simulations were performed on SuperMUC
at LRZ-Munich, on Datura at AEI-Potsdam and on LOEWE at CSC-Frankfurt.

%%%%%%%%%%%%%%%%%%%%%%%%%%%%%%%%%%%%%%%%%%%%%%%%%%%%%%%%%%%%%%%%%
%%%   REFERENCES
%%%%%%%%%%%%%%%%%%%%%%%%%%%%%%%%%%%%%%%%%%%%%%%%%%%%%%%%%%%%%%%%%

\bibliographystyle{apsrev4-1-noeprint}
% \bibliography{aeireferences}

%%%%%%%%%%%%%%%%%%%%%%%%%%%%%%%%%%%%%%%%%%%%%%%%%%%%%%%%%%%%%%%%%
%%%%%%%%%%%%%%%%%%%%%%%%%%%%%%%%%%%%%%%%%%%%%%%%%%%%%%%%%%%%%%%%%

\end{document}